\documentclass[%
 aip,
 amsmath,amssymb,
 reprint,%
]{revtex4-1}

\usepackage{graphicx}
\usepackage{dcolumn}
\usepackage{bm}
\usepackage[table,xcdraw]{xcolor}
\usepackage[utf8]{inputenc}
\usepackage[T1]{fontenc}
\usepackage{mathptmx}
\begin{document}

\preprint{AIP/123-QED}

\title[3D Flow Field Measurements Outside Nanopores]{3D Flow Field Measurements Outside Nanopores}

\author{Jeffrey Mc Hugh}
 \affiliation{Cavendish Laboratory, University of Cambridge, Cambridge CB3 0HE, United Kingdom}
\author{Alice L. Thorneywork}
 \affiliation{Cavendish Laboratory, University of Cambridge, Cambridge CB3 0HE, United Kingdom}
\author{Kurt Andresen}
 \email[Email: ]{kandrese@gettysburg.edu}
 \affiliation{Department of Physics, Gettysburg College, Gettysburg, PA 17325, United States of America}
\author{Ulrich F. Keyser}
 \email[Email: ]{ufk20@cam.ac.uk}
 \affiliation{Cavendish Laboratory, University of Cambridge, Cambridge CB3 0HE, United Kingdom}

\date{\today}

\begin{abstract}
We demonstrate a non-stereoscopic, video-based particle tracking system with optical tweezers to study fluid flow in 3D in the vicinity of glass nanopores. In particular, we used the Quadrant Interpolation algorithm to extend our video-based particle tracking to displacements out of the trapping plane of the tweezers. This permitted the study of flow from nanopores oriented at an angle to the trapping plane, enabling the mounting of nanopores on a micromanipulator with which it was then possible to automate the mapping procedure. Mapping of voltage driven flow in 3D volumes outside nanopores revealed polarity dependent flow fields. This is in agreement with the model of voltage driven flow in conical nanopores depending on the interaction of distinct flows within the nanopore and along the outer walls.
\end{abstract}

\maketitle

\begin{figure}
\includegraphics{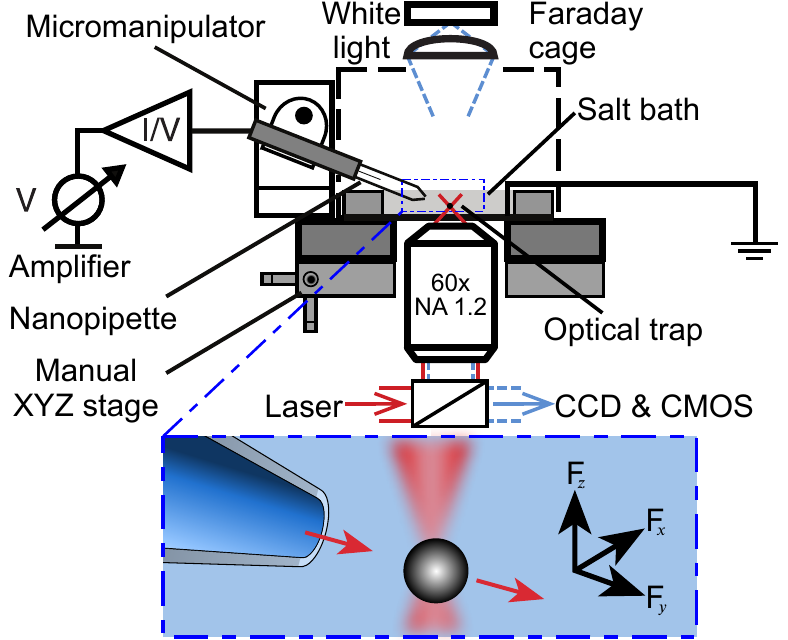}
\caption{Schematic of the experimental setup showing relative positions of glass nanopore, micromanipulator and salt bath. An illustration of the movement of a bead resulting from electroosmotic flow due to an applied voltage is shown in the blue dashed box. The force on the bead can have components in the $x$, $y$ and $z$ directions depending on the position of the pore relative to the bead.}
\label{fig:fig1}
\end{figure}

\begin{figure}
\includegraphics{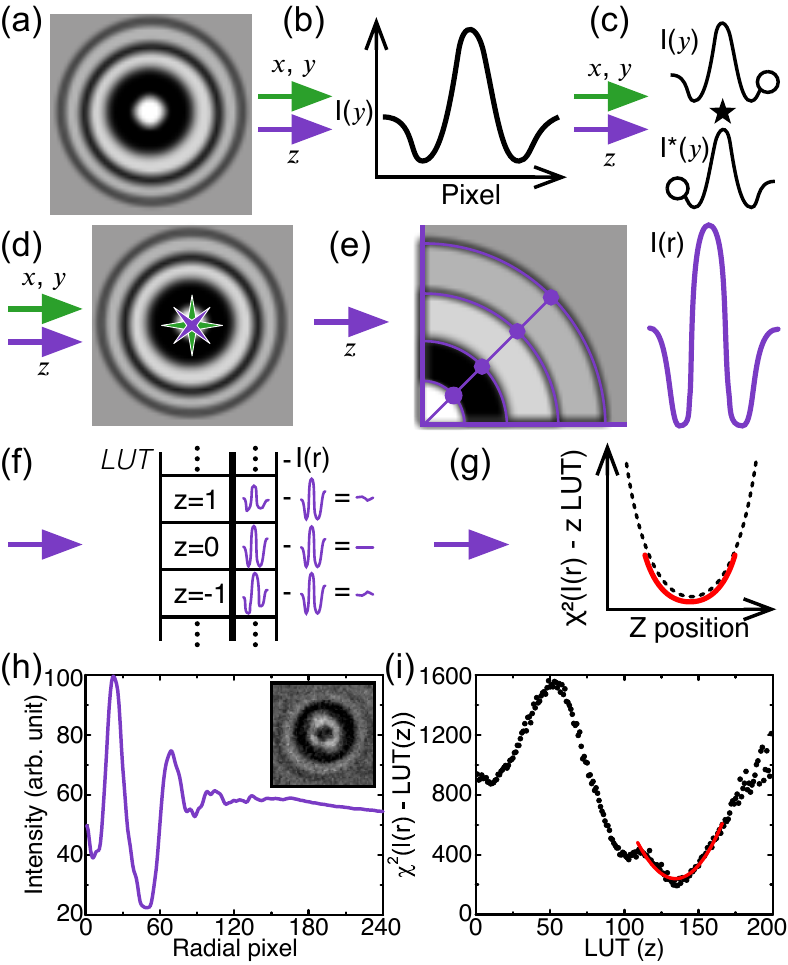}
\caption{3D particle tracking. Both in-plane ($x$, $y$) tracking and out-of-plane ($z$) tracking require identification of the particle center (a)--(d), while (e)--(g) are specific to $z$ tracking. (a) A frame is cropped to a $20\times20$ px$^2$ ($x$, $y$) or $80\times{}80$ px$^2$ ($z$) sub-array approximately centered on the trapped bead. (b) Intensity profiles in $x$ and $y$ are obtained by averaging over $4\times20$ px$^2$ rectangular bins, producing I($x$) and I($y$). (c) The cross-correlation between each intensity profile and its reverse, I$^{\star}(x)$, I$^{\star}(y)$, is found. (d) A quadratic fit to the cross-correlation determines the particle center ($x$, $y$ position) with sub-pixel accuracy. To obtain the $z$ coordinate: (e) The $80\times{}80$ px$^2$ frame is quartered and a radial intensity profile is calculated per quadrant. (f) The 4 profiles are averaged and this mean profile is compared to each entry of a lookup table (LUT). (g) $\chi^2$ difference is used to compare the profile to the LUT, and a quadratic fit around the minimum of the $\chi^2$ difference enables sub-LUT resolution. (h) Typical intensity profile of a trapped particle (inset $80\times{}80$ px$^2$) averaged from 20 radial profiles over one quadrant. (i) $\chi^{2}$ difference LUT comparison for the shown radial profile, with quadratic fit shown in red.}
\label{fig:fig2}
\end{figure}

\begin{figure*}
\includegraphics[width=\textwidth]{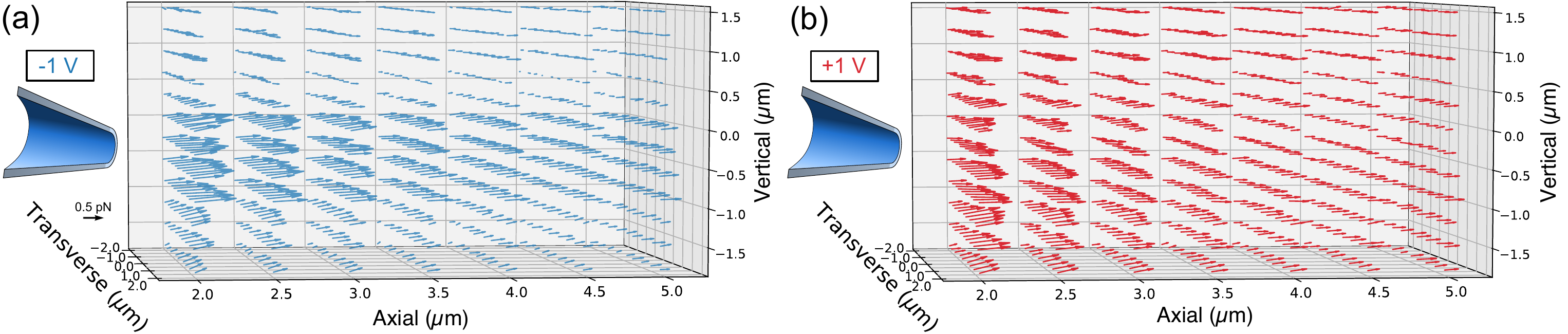}
\caption{3D vector plots of the voltage induced flow fields in the vicinity of a nanopore in 10 mM NaCl solution. (a) Flow in the volume outside a nanopore when -1 V is applied. (b) Flow in the same volume when +1 V is applied. Axes are labelled relative to the nanopore; axial is parallel to the long axis of the capillary, transverse is perpendicular to that and the vertical axis is orthogonal to the transverse--axial ($x$, $y$) plane. Arrow length corresponds to the magnitude of the force experienced by the trapped bead as a result of the flow at a given position.}
\label{fig:fig3}
\end{figure*}

Solid-state nanopores underpin a simple but powerful molecular sensing platform that has blossomed into an important tool for biophysics and beyond. The basic operating principle of these sensors involves using an electric field to drive analytes through the nanopore, with the translocation observed as a change in ionic current. The operation of nanopore devices is, however, governed by a subtle interplay of phenomena: target molecules are driven electrophoretically~\cite{Mathe2004, Keyser2006, Laohakunakorn2015b}, and this driving force is opposed by a viscous drag force~\cite{VanDorp2009}. This drag force arises not only from Stokes drag, but also from counterions on the target molecule driven in the opposite direction by the electric field~\cite{VanDorp2009} and from the voltage-driven, electroosmotic flow (EOF) from the pore surface~\cite{Laohakunakorn2015b, Ghosal2019}. Characterising EOF and broader nanopore flow behaviour offers novel physical insights into the hydrodynamics of nanochannel systems~\cite{Schoch2008} and has applications in molecular sensing~\cite{Huang2017}, gel electrophoresis~\cite{VanDorp2009} and electrophysiology~\cite{Rabinowitz2019}. At these small length scales, however, the relevant forces have pico-Newton magnitudes, leading to significant challenges in their measurement.

Optical tweezers, with sensitivity to forces on the order of 1~pN, can be used to measure the forces of these nanoscale fluid flows and map flow fields~\cite{Laohakunakorn2015b}, and have previously revealed salt dependent flow behaviors around glass capillary nanopores~\cite{Laohakunakorn2015b, McHugh2019}. The graphic in the blue dashed box below the schematic in Fig.~\ref{fig:fig1} illustrates the principle of measuring nanopore electroosmotic flow using optical tweezers. Specifically, a $2.16$~$\mu$m diameter polystyrene bead (Kisker, Germany) suspended in salt solution is held in an optical trap just outside the pore entrance. Applying a voltage causes fluid flow around the nanopore that perturbs the trapped particle and displaces it in the direction of the flow, thereby revealing information about the field's direction and strength. By changing the position of the nanopore tip relative to the trapped bead while recording the forces, the flow field is mapped. 

Previous studies considered flows through nanopores in the $x$, $y$ trapping plane of the optical tweezers, by using in-plane tracking of trapped particles to measure flow forces in the $x$, $y$ plane. Full characterisation of flow profiles also requires mapping of the out-of-plane $z$ component of the flows, orthogonal to the tracking plane. As such, here, we outline an extension of the flow mapping technique to flow forces in 3D.

To achieve this, glass capillary nanopores were mounted on a micromanipulator at an angle to the trapping plane, enabling straightforward positioning of the pore relative to a trapped bead (Fig.~\ref{fig:fig1}). This arrangement permits automated mapping of flow forces in the 3D space around the nanopore. Mounting the nanopore on a micromanipulator allows for open bath nanopore experiments, making the changing of solution or addition of analytes possible and emulates electrophysiology experimental setups. Thus, this platform offers an opportunity to study flows in electrophysiology experiments with glass nanopores, such as the concentration gradient driven outflows believed to prevent nanopore clogging during in-vivo experiments~\cite{Jayant2019, Rabinowitz2019}.

Fig.~\ref{fig:fig1} shows a schematic of the setup used to measure nanopore flow-fields in 3D. Nanopores have a nominal diameter of 160~nm and are produced using a laser pipette puller (P2000/F, Sutter Instruments, USA) from quartz glass capillaries. Nanopores are assembled into a HL-U pipette holder (Molecular Devices, USA). The holder is connected to a patch-clamp amplifier (Axon Axopatch 200B, Molecular Devices, USA) via the headstage. The headstage is itself mounted on a micromanipulator (Patchstar, Scientifica, UK). The micromanipulator enables the positioning of the nanopore in 3D space, with a resolution of 20~nm and a travel of 20~mm in $x$, $y$ and $z$. The headstage and nanopore are enclosed by a Faraday cage during experiments to minimise both electrical noise and mechanical noise from air flow. The amplifier outputs to a control PC using a PCIe-6251 data acquisition card (National Instruments, USA), capable of 1.25 million samples per second. This allows the amplifier to be controlled and recorded from using a custom written LabVIEW program (LabVIEW 2016, National Instruments, USA). The micromanipulator is managed with the same program, enabling the scanning of the nanopipette through a range of $x$, $y$ and $z$ positions automatically over the course of an experiment. The optical tweezers consists of a $1064$~nm laser focused through a $60\times$ water-immersion objective with $1.2$~NA in an inverted microscope configuration. Further details of the optics can be found here~\cite{Otto2010}.

Forces are measured through video-based tracking of the trapped particles. To track particle motion in 3D, both in-plane ($x$, $y$) and out-of-plane ($z$) tracking are performed separately, with $x$, $y$ tracking performed live and $z$ tracking performed post-acquisition. The $x$, $y$ tracking of a trapped particle is performed using a CMOS camera (MC1362, Mikrotron, Germany). The CMOS camera is interfaced to the primary computer using a PCIe-1433 frame grabber with a full configuration (two cables) Camera Link (National Instruments, USA). To achieve adequate sampling for particle tracking, the camera records at $1000$~fps. Images are taken over a region of interest (ROI) of $128\times100$~px$^2$ centered to the optical trap.

To initiate the in-plane tracking algorithm in LabVIEW, a cursor is used to specify the approximate bead center. The algorithm then follows the steps detailed in Fig.~\ref{fig:fig2} (a)--(d). Initially, a sub-array of the ROI centered on the bead ($20\times20$~px$^2$ in size) is defined (Fig.~\ref{fig:fig2} (a)). Mean intensity profiles in $x$ and $y$, I($x$) and I($y$), are obtained by averaging over $4\times20$~px$^2$ rectangular bins centered to the sub-array in $x$ and $y$ respectively (Fig.~\ref{fig:fig2} (b)). These profiles are reversed to create I$^{\star}(x)$ and I$^{\star}(y)$, and the cross-correlation between each profile and its reverse is performed (Fig.~\ref{fig:fig2} (c)), e.g. I$(x)\bigstar$I$^{\star}(x)$. A quadratic fit to the 7 central points of the cross-correlation is used to determine the center position with sub-pixel accuracy (Fig.~\ref{fig:fig2} (d)).

To track the trapped particle's movement out-of-plane, the particle is simultaneously imaged using a CCD camera (DMK31AF03, Imaging Source, Germany) at 30 fps and $1024\times768$ px$^2$, with synchronised recording to the CMOS controlled with LabVIEW. The routine is based on the Quadrant Interpolation algorithm developed by van Loenhout et al.~\cite{VanLoenhout2012} and is again illustrated in Fig.~\ref{fig:fig2}. Center finding (Fig.~\ref{fig:fig2} (a)--(d)) follows the same method as $x$, $y$ tracking, with some differences arising from the lower bit depth of the CCD camera noted here. In particular, the median intensity of the edges of the ROI in each dimension is used for initial background subtraction to improve the signal-to-noise ratio. The pixel intensity in each dimension is then summed and divided by the total number of pixels to estimate the particle center. The ROI is cropped to a sub-array of $80\times{}80$ px$^2$ centered on that estimate and the cross-correlation is computed and fit with a quadratic as before to determine the center with sub-pixel accuracy.

Using this more accurate estimate of the center, the cropped image is quartered and a radial coordinate system is imposed on each quadrant. A radial intensity profile is calculated per quadrant, I$(r)$ (Fig.~\ref{fig:fig2} (e)) and the four I$(r)$ are averaged together to obtain a mean radial intensity profile for that frame. The mean profile is compared to a series of intensity profiles in a lookup table (LUT), where each profile corresponds to a specific $z$-position (Fig.~\ref{fig:fig2} (f)). LUT profiles are created by applying the technique described so far (Fig.~\ref{fig:fig2} (a)--(e)) to a particle fixed on a coverslip. A piezo stage is used to move in controlled $z$ steps, with the step size dictating the resolution of the LUT. By fitting a quadratic about the minimum of the $\chi^2$ difference between the measured profile and LUT profiles, the $z$ position of the trapped bead is determined with sub-LUT accuracy (Fig.~\ref{fig:fig2} (g)). A typical mean intensity profile generated by averaging 20 profiles from one quadrant is shown in Fig.~\ref{fig:fig2} (h). The $\chi^2$ comparison of this profile to the LUT is shown in Fig.~\ref{fig:fig2} (i) with the quadratic fit overlaid. We determined the resolution of our $z$-tracking implementation to be 50~nm.

Combining 3D tracking with a micromanipulator mounted nanopore, we recorded the flow fields generated by EOF in a volume outside a glass nanopore. As mentioned earlier, this recording procedure is automated; at each position in the volume a negative voltage is applied, followed by a positive voltage, while the position of the bead is recorded. After this the micromanipulator moves the nanopore to a new position in the volume and the procedure is repeated. Fig.~\ref{fig:fig3} shows a typical map generated from a 160~nm pore in 10~mM NaCl solution. In Fig.~\ref{fig:fig3} (a), at -1~V a stronger central flow can be seen with weaker flow in the spaces above and below. In contrast, in Fig.~\ref{fig:fig3} (b) at +1~V, a more uniform flow field is observed. That the two flow fields are not mirror images of each other corroborates the inner/outer flow model of conical glass nanopore EOF developed by Laohakunakorn et al.~\cite{Laohakunakorn2015b}, whereby bulk flow is the result of the interplay of opposing flows through the nanopore and along the outer walls. The vertical component of individual flow forces is strongest and thus most visible close to the tip. Though the glass nanopore was inserted into the salt bath at an angle of approximately 15$^{\circ}$, the flow forces are not directed at this angle, demonstrating the non-ideal geometric reality at the tip of glass nanopores.

In this Note we have demonstrated how an optical tweezers equipped for 2D particle tracking can be adapted to perform 3D tracking, with two cameras for $x, y$ and $z$ tracking respectively. By combining this with a micromanipulator mounted nanopore, we have developed an automated 3D flow mapping platform. Measuring flow forces at each point in 3D provides a means to observe the true geometry of a glass nanopore aperture, we believe this offers a route to quantify their geometry more precisely. The use of a 30 fps camera makes our implementation of $z$-tracking inexpensive and accessible. While the low sampling rate precludes study of out-of-plane forces using variance and power spectrum based analyses, a higher frame-rate camera could be introduced if needed. As the off-line implementation of $z$-tracking is hardware independent, it is readily portable to other experiments needing 3D particle tracking.
\vspace{-6.5mm}
\begin{acknowledgments}
\vspace{-4mm}
J. Mc H. acknowledges funding from AFOSR (Grant No. FA9550-17-1-0118). A. L. T. acknowledges support from the University of Cambridge Ernest Oppenheimer Fund. U. F. K. is supported by ERC Consolidator Grant DesignerPores 647144.
\end{acknowledgments}

\bibliography{refs}

\end{document}